\documentclass[preprint,5p,times,twocolumn, compress, sort]{elsarticle}

\usepackage{amssymb}
\usepackage{siunitx}
\ifdefined\unit\else
  \ifdefined\NewCommandCopy
    \NewCommandCopy\unit\si
  \else
    \NewDocumentCommand\unit{O{}m}{\si[#1]{#2}}
  \fi
\fi
\usepackage{hyperref}
\hypersetup{
   colorlinks=true,
   linkcolor=blue,
   filecolor=magenta,      
   urlcolor=cyan,
}

\journal{NIMA}

\begin{document}

\begin{frontmatter}



\title{Defect characterization studies on neutron irradiated boron-doped silicon pad diodes and Low Gain Avalanche Detectors}


\author[1]{Anja Himmerlich\corref{cor1}}
\ead{anja.himmerlich@cern.ch}
\author[1]{Nuria Castello-Mor}
\author[1]{Esteban Curr\'{a}s Rivera}
\author[{1}]{Yana Gurimskaya}
\author[1,3]{Vendula Maulerova-Subert}
\author[1]{Michael Moll}
\author[2]{Ioana Pintilie}
\author[3]{Eckhart Fretwurst}
\author[3]{Chuan Liao}
\author[3]{Jörn Schwandt}

\cortext[cor1]{Corresponding author}
\address[1]{European Organization for Nuclear Research, CERN, Esplanade des Particules 1, Geneva, 1211, Switzerland}
\address[2]{National Institute of Materials Physics, NIMP, Str. Atomistilor 105 bis, Bucharest, RO-77125, Romania}
\address[3]{Institute for Experimental Physics, University of Hamburg, Luruper Chaussee 149, Hamburg, 22761, Germany}
%%\affiliation[1]{organization={European Organization for Nuclear Research, CERN},%Department and Organization
%%            addressline={Esplanade des Particules 1},
%%            city={Geneva},
%%            postcode={1211}, 
%%            country={Switzerland}}
%%
%%\affiliation[2]{organization={National Institute of Materials Physics, NIMP},%Department and Organization
%%            addressline={Str. Atomistilor 105 bis}, 
%%            city={Bucharest},
%%            postcode={RO-77125}, 
%%            country={Romania}}
%%
%%\affiliation[3]{organization={Institute for Experimental Physics, University of Hamburg},%Department and Organization
%%            addressline={Luruper Chaussee 149}, 
%%            city={Hamburg},
%%            postcode={22761}, 
%%            country={Germany}}

\begin{abstract}
High-energy physics detectors, like Low Gain Avalanche Detectors (LGADs) that will be used as fast timing detectors in the High Luminosity LHC experiments, have to exhibit a significant radiation tolerance. Thereby the impact of radiation on the highly boron-doped gain layer that enables the internal charge multiplication, is of special interest, since due to the so-called Acceptor Removal Effect (ARE) a radiation-induced deactivation of active boron dopants takes place. In this paper we present defect-spectroscopy measurements (Deep-Level Transient Spectroscopy and Thermally Stimulated Current technique) on neutron irradiated p-type silicon pad diodes of different resistivity as well as LGADs irradiated at fluences up to \num{1e15},\unit{n_{eq}\per\centi\square\meter}. Thereby we show that while for the silicon pad diodes irradiated with electrons, neutrons or protons the determination of defect electronic properties and defect introduction rates is straightforward, DLTS and TSC measurements on LGADs are strongly influenced by the impact of the gain layer. It is shown that the measurability of the capacitance of the gain layer shows a strong frequency and temperature dependence leading to a capacitance drop in DLTS and non-reliable measurement results. With TSC defects formed in the LGADs can be very nicely observed and compared to the defects formed in the silicon pad diodes. However the exact assignment of defects to the gain layer or bulk region remains challenging and the charge amplification effect of the LGADs impacts the exact determination of defect concentrations.  Additionally, we will demonstrate that depending on the TSC measurement conditions defect induced residual internal electric fields are built up in the irradiated LGADs that are influencing the current signal of carriers emitted from the defect states.  
\end{abstract}


%\item Defect spectroscopy studies (DLTS and TSC) on neutron irradiated LGADs are compared to results obtained from measurements on irradiated silicon pad diodes. 
%\item Defect spectroscopy studies on LGADs are significantly influenced by the highly doped LGAD gain layer, making the exact determination of defect characteristics challenging. 
%\item The measured gain layer capacitance shows a strong frequency and temperature dependence.
%\item Defect induced internal electrical fields can built up in irradiated LGADs. 
%\item Experimentally determined B$_\text{i}$O$_\text{i}$ introduction rates for silicon pad diodes irradiated with neutrons, electrons or protons are given. 


\end{frontmatter}

%\linenumbers

%% main text %%%%%%%%%%%%%%%%%%%%%%%%%%%%%%%%%%%%%%%%%%%%%%%%%%%%%%%%%%%%%%%%%%%%%%%%%%%%%%%%%%%%%%%%%%%%%%%%%%%%%%%%%%%%%%%%%%%%%%%%%%%
\section{Introduction}
\label{sec:introduction}
Low Gain Avalanche Detectors (LGADs) are characterized by their high precision timing performance and are the selected technology for the Atlas High-Granularity Timing Detector (HGTD) as well as the CMS Endcap timing layer (ETL) \cite{CERN-LHCC-2017-027,CERN-LHCC-2018-023}. The operation of such sensors in the HL-LHC experiments requires a high radiation tolerance up to a 1\,MeV neutron equivalent fluence $\Phi_\text{eq}$ of about \num{2e16}\,\unit{\per\centi\square\meter}. 
For LGADs a radiation induced degradation in the device performance can be observed that correlates with changes in the effective doping concentration $N_\text{eff}$ of the highly doped gain layer \cite{Kramberger2015JIST}. Normally the doping of such layer is up to \num{1e17}\,cm$^{-3}$ and enables charge multiplication due to impact ionization. The degradation becomes evident in a decrease in the signal gain with increasing particle fluence resulting in a disappearance of the multiplication effect at fluences of about \num{2e15}cm$^\text{-2}$ \cite{Ferrero2019NIMA}. 

The radiation-induced deactivation of boron in si\-li\-con (Si) is well-known as so-called Acceptor Removal Effect (ARE). Thereby, due to the interaction with high-energy particles, Si atoms are released from their lattice site and become Si interstitials (Si$_\text{i}$) which are very mobile, even at low temperatures, and interact via the Watkins replacement mechanism preferentially with boron and carbon atoms \cite{Watkins2000, Jones09} forming boron and carbon interstitials (B$_\text{i}$, C$_\text{i}$). These interstitials can further interact and create boron and carbon related defects like e.g. B$_\text{i}$B$_\text{s}$, B$_\text{i}$O$_\text{i}$, B$_\text{i}$C$_\text{s}$, C$_\text{i}$C$_\text{s}$ or C$_\text{i}$O$_\text{i}$  \cite{Kimerling1989, MOLL2019Vertex}. Thereby the interstitial boron - interstitial oxygen complex (B$_\text{i}$O$_\text{i}$) is generally considered as main responsible defect for the boron deactivation. Although recent publications also state a B$_\text{Si}$Si$_\text{i}$ as possible defect structure to explain the ARE \cite{Lauer2020PSSA}, in this publication we will follow the so far widely accepted assumption of a B$_\text{i}$O$_\text{i}$ defect structure. Its creation is coupled with the deactivation of one negatively charged boron atom B$_\text{s}$ and the formation of a donor type defect with an energy level in the upper part of the Si band gap (E$_\text{C}$\,-\,0.25\,eV). In this regard, the B$_\text{i}$O$_\text{i}$ formation contributes with a factor of two to the changes of the space charge in the depletion region \cite{MOLL2019Vertex}. The formation of B$_\text{i}$O$_\text{i}$ competes with the formation of C$_\text{i}$O$_\text{i}$ which induces a hole trap at E$_\text{V}$\,+\,0.36\,eV and does not contribute to the ARE. This competition gives an explanation of the improved radiation hardness of carbonated LGAD gain layers \cite{Ferrero2019NIMA}. In summary, the described defect kinetic model assumes that all interstitials created during the radiation interaction are forming either B$_\text{i}$O$_\text{i}$ or C$_\text{i}$O$_\text{i}$ defects \cite{MOLL2019Vertex}. This is very nicely reproduced in the experimentally observed dependency of B$_\text{i}$O$_\text{i}$ introduction rates (IR = (defect concentration)/(fluence)) on the initial boron doping concentration for p-type silicon devices irradiated with fluences of up to 10$^{15}$\,\unit{n_{eq}\per\centi\square\meter} \cite{MOLL2019Vertex}. However, it seems not be valid anymore for highly doped silicon, like the gain layers of LGADs, irradiated at fluences $>$\,10$^{15}$\,\unit{n_{eq}\per\centi\square\meter} \cite{MOLL2019Vertex}, since the generation rates that reflect the observed deactivation of boron extracted by changes of the macroscopic device properties are much higher as expected from the defect kinetic model \cite{Ferrero2019NIMA, MOLL2019Vertex}. That raises the question if other defect structures might lead or contribute to the deactivation of acceptors in the highly doped LGAD gain layers. In order to trace back this question we present defect spectroscopy studies on silicon pad diodes and LGADs using the Deep-level Transient Spectroscopy (DLTS) and Thermally Stimulated Current (TSC) techniques.   
%-------------------------------------
\section{Experimentals}
\label{sec:experimental}
The measurements were performed on LGADs and PIN diodes from CNM (Centro Nacional de Mi\-cro\-e\-lec\-tró\-ni\-ca, Barcelona, Spain) and HPK (Hamamatsu Photonics, Ja\-pan). The LGAD structure is $n^{++}$-$p^{+}$-$p$-$p^{++}$ with a thin highly boron-doped multiplication layer ($p^{+}$) of around 2\,\unit{\micro\meter} (this layer is missing in the PIN diodes), a low doped active bulk region of about 50\,\unit{\micro\meter}  ($p$) as well as the highly doped electrodes ($n^{++}$ and $p^{++}$). 
 Irradiation was performed with reactor neutrons at the JSI in Ljubljana (Slovenia). The sample overview as well as the irradiation fluences are given in Table\,\ref{tab:samples}. The fluences given in this paper are normalized to 1 MeV neutron equivalent values ($\Phi_\text{eq}$) using the non-ionizing energy loss scaling (NIEL). The LGADs from CNM are from Run 11486 with an active area of 0.09\,cm$^{2}$ and a physical thickness of 351\,\unit{\micro\meter}. They consist of a low resistivity p-type support wafer and a 50\,\unit{\micro\meter} boron-doped active layer with a resistivity of about 5\,k$\Omega$cm into which, during the sensor processing, an active p-type multiplication layer is implanted underneath the front electrode. The HPK LGADs have an active area of 0.0169\,cm$^{2}$ and consists of an a 300\,\unit{\micro\meter} support wafer, a 50\,\unit{\micro\meter} active p-type layer and an active highly-boron doped p-type multiplication layer. After irradiation all samples were annealed by default 10\,min at 60\unit{\celsius}.\\
The defect spectroscopy measurements on the LGADs and PIN diodes are, among others, compared to similar studies performed on single boron-doped $n^{+}$-$p$-$p^{+}$ silicon pad diodes produced at CiS (Forschungsinstitut für Mikrosensorik GmbH, Erfurt, Germany)  \cite{Gurimskaya2020}. These diodes consist of an epitaxial grown boron-doped bulk layer of 50\,\unit{\micro\meter} that vary in resistivity from 10\,$\Omega$cm to 1\,k$\Omega$cm. The active area of those devices is \num{6.927E-2}\,cm$^{2}$. They were irradiated with neutrons at JSI \cite{JSI}, with 24\,GeV/c protons at IRRAD proton facility (CERN) \cite{PS}, with 230\,MeV protons at Boston General Hospital (USA) or with 200\,MeV electrons at CLEAR (CERN)  \cite{CLEAR}, and annealed afterwards for 10\,min at 60\unit{\celsius}. As hardness factors for the $\Phi_\text{eq}$ calculation 0.62 was used for 23\,GeV proton irradiation, 0.95 for 230\,MeV proton irradiation and 0.082 for 200\,MeV electron irradiation. \\
In order to investigate the macroscopic properties of the non-irradiated and irradiated diodes Capacitance-Voltage ($C$-$V$) measurements with different frequencies as well as Current-Voltage ($I$-$V$) measurements were performed. As defect spectroscopy methods Deep Level Transient Spectroscopy (DLTS) and Thermally Stimulated Current technique (TSC) were applied. 
For DLTS a commercial system from PhysTech GmbH \cite{Phystech} was used. Thereby capacitance transients were analysed that were recorded after charge carrier injection at temperatures in the range from 20\,K to 280\,K, using three different time windows (20\,ms, 200\,ms and 2\,s). The injection pulse was varied in time (t$_\text{p}$\,=\,0.1\,--\,100\,ms) and pulse voltage (UP). To inject only majority carriers a pulse voltage of UP\,=\,-\,0.6\,V was chosen, while for minority and majority carrier injection the pulse voltage was set to +\,2\,V. Before and after the injection pulse the device was put under reverse bias UR (typically UR\,=\,-10\,V). The measurement frequency was 1\,MHz. \\
TSC measurements were performed in the temperature range from 20\,K to 220\,K by using a Keithley electrometer and a Labview based DAQ. The typical measurement cycle for spectra presented in this paper consists of three steps:\\
(1) Cooling down: A reverse bias \unit{UR_{down}} is applied to the diode at high temperature ($\geq$\,220\,K). Afterwards the biased diode is subsequently cooled down to a certain filling temperature \unit{T_{fill}} (20\,K to 90\,K). This step assures the release of charges and the availability of unoccupied defect states. \\
(2) Filling step: At \unit{T_{fill}} a filling pulse UP (20\,V) is applied to the sensor during a specific filling time (\unit{t_{fill}}\,=\,60\,s\,--\, 360\,s). During this steps the defect states are occupied by majority and/or minority carriers.\\
(3) Heating up: the diode is put back under a reverse bias \unit{UR_{up}} and the temperature is raised with a constant heating rate of 11\,K/min from \unit{T_{fill}} to 220\,K. During this time the current signal induced by thermal emission of carriers from the defect levels is recorded. In case that \unit{UR_{down}} is equal to \unit{UR_{up}} the reverse bias is just named UR in this paper.   \\
From the recorded TSC spectra the defect concentrations $N_\text{t}$ were determined by integration over the observed TSC peaks $Q_\text{t}$ using the following equation \cite{Moll-thesis}:
\begin{equation}
    N_{t} = 2 \frac{Q_{t}}{q_{0}Ad}
    \label{eq:Nt}
\end{equation}
with $q_{0}$: the elementary charge as well as $A$: the area and $d$: the thickness of the active region.
%%%%%%%%%% Figure %%%%%%%%%%%%%%%%%%%%%%%%%%%%%%
\begin{figure}[htb]
    \centering
    \includegraphics[width=1\columnwidth]{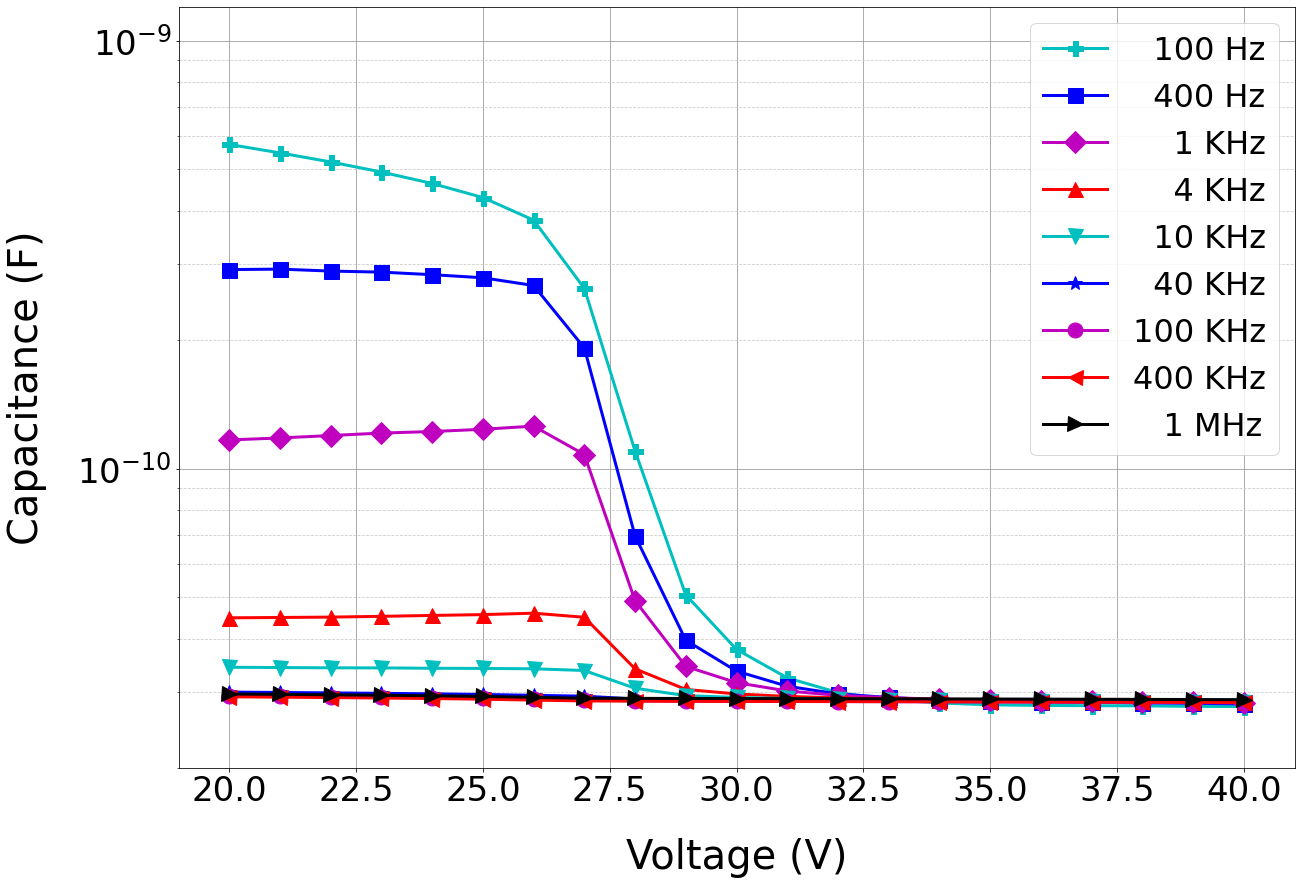}
    \caption{$C$-$V$ measurements performed on a CNM LGAD, neutron irradiated with \num{1e14}\,\unit{n_{eq}\per\centi\square\meter}. The measurements were performed at a temperature of 20$^\circ$C at different frequencies from 100\,Hz up to 1\,MHz.}
    \label{fig:CV}
\end{figure}
%%%%%%%%%%%%%%%%%%%%%%%%%%%%%%%%%%%%%%%%%%%%%%%%
%%%%%%%%%% Figure %%%%%%%%%%%%%%%%%%%%%%%%%%%%%%
\begin{figure}[htb]
    \centering
    \includegraphics[width=1\columnwidth]{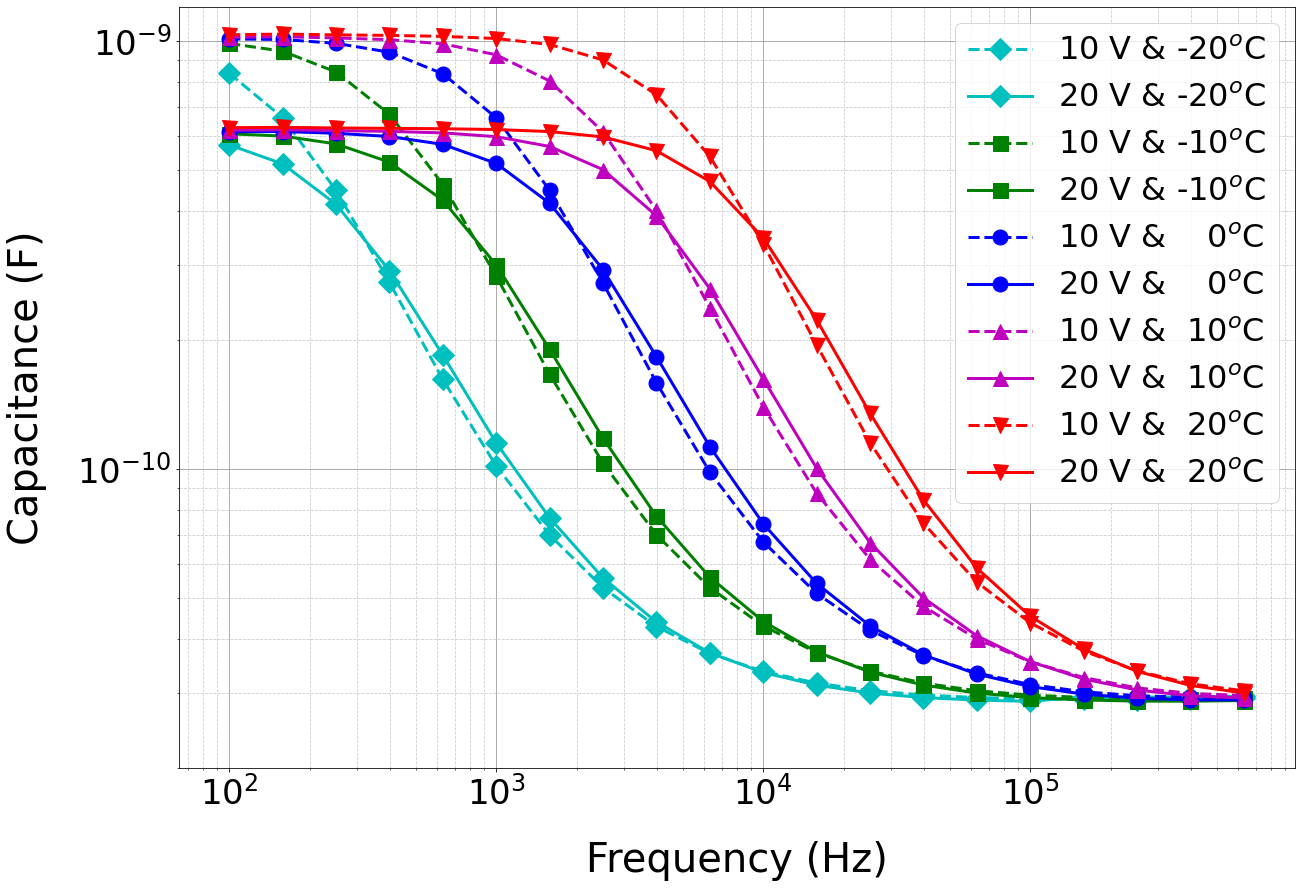}
    \caption{Capacitance measured at certain bias voltages and temperatures in dependency of the frequency. The measurement were performed on a CNM LGAD neutron irradiated with \num{1e14}\,\unit{n_{eq}\per\centi\square\meter}. }
    \label{fig:CF}
\end{figure}
%%%%%%%%%%%%%%%%%%%%%%%%%%%%%%%%%%%%%%%%%%%%%%%%
\begin{table*}[h!] 
\centering
\begin{tabular}{l c c c c}
\hline
 & & $\Phi_\text{eq}$\,(cm$^\text{-2}$) & $V_\text{depl}$\,(V) & $V_\text{GL}$\,(V)  \\
\hline
LGAD (HPK) & W36 S3-L15P5 & \num{1e13} & - & 50.8 (51.2) \\
PIN (HPK) & W42 S4-L14P5 & \num{1e13} & 6.5 (7.2) & - \\
\hline
LGAD (CNM) & r11486 W2-U23 & \num{1e14} & - &  27.0 (39.0)  \\
LGAD (CNM) & r11486 W3-A12 & \num{1e14} & - & 26.0 (30.5) \\
PIN (CNM) & r11486 W2-X22 & \num{1e14} & 0.6 (1.8) &  -\\
\hline
LGAD (CNM) & r11486 W2-W22 & \num{1e15} &  - &  14.0 (42.5) \\
\hline
\end{tabular}
\caption{Samples overview: The samples were neutron irradiated according to the given fluences $\Phi_\text{eq}$. For the LGADs the gain layer depletion voltages $V_\text{GL}$ and for the PIN diodes the full depletion voltages $V_\text{depl}$ are given in this table. The values were extracted from $C$-$V$ measurements performed at -\,20$^{\circ}$C and with 10\,kHz measurement frequency. In brackets the values of the unirradated sensors are added.}
\label{tab:samples}
\end{table*}
%%%%%%%%%% Figure %%%%%%%%%%%%%%%%%%%%%%%%%%%%%%
\begin{figure}[h!]
    \includegraphics[width=1\columnwidth]{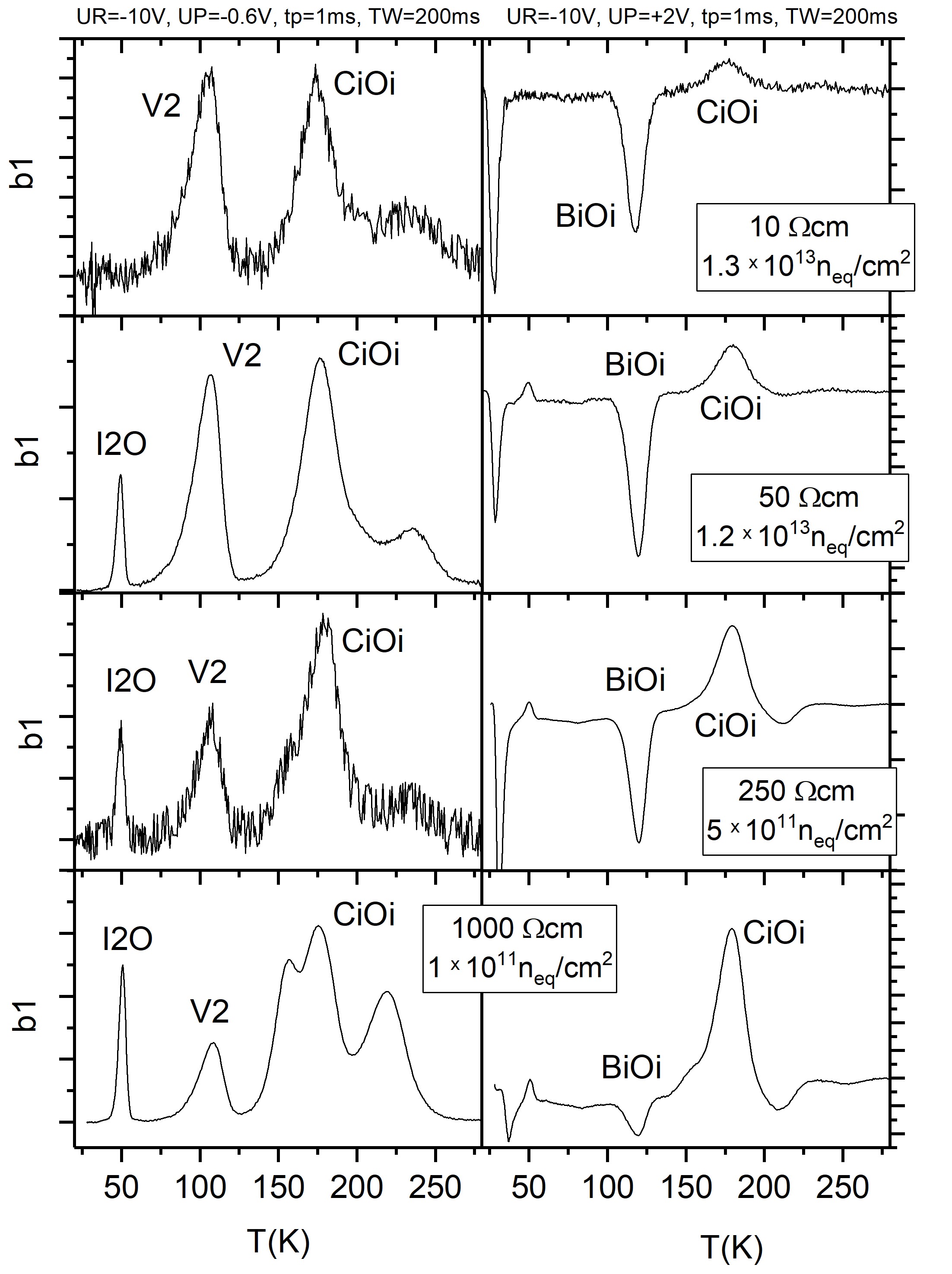}
    \caption{DLTS spectra of EPI silicon pad diodes. The resistivity of the diodes vary due to the different boron content from 10\,$\Omega$cm to 1\,k$\Omega$cm. The spectra on the left side were obtained when injecting majority carriers during the pulse step, while the spectra on the right side were recorded after majority and minority carrier injection.}
    \label{fig:DLTS-EPI}
\end{figure}
%%%%%%%%%%%%%%%%%%%%%%%%%%%%%%%%%%%%%%%%%%%%%%%%
%%%%%%%%%% Figure %%%%%%%%%%%%%%%%%%%%%%%%%%%%%%
\begin{figure}[htb]
    \centering
    \includegraphics[width=1\columnwidth]{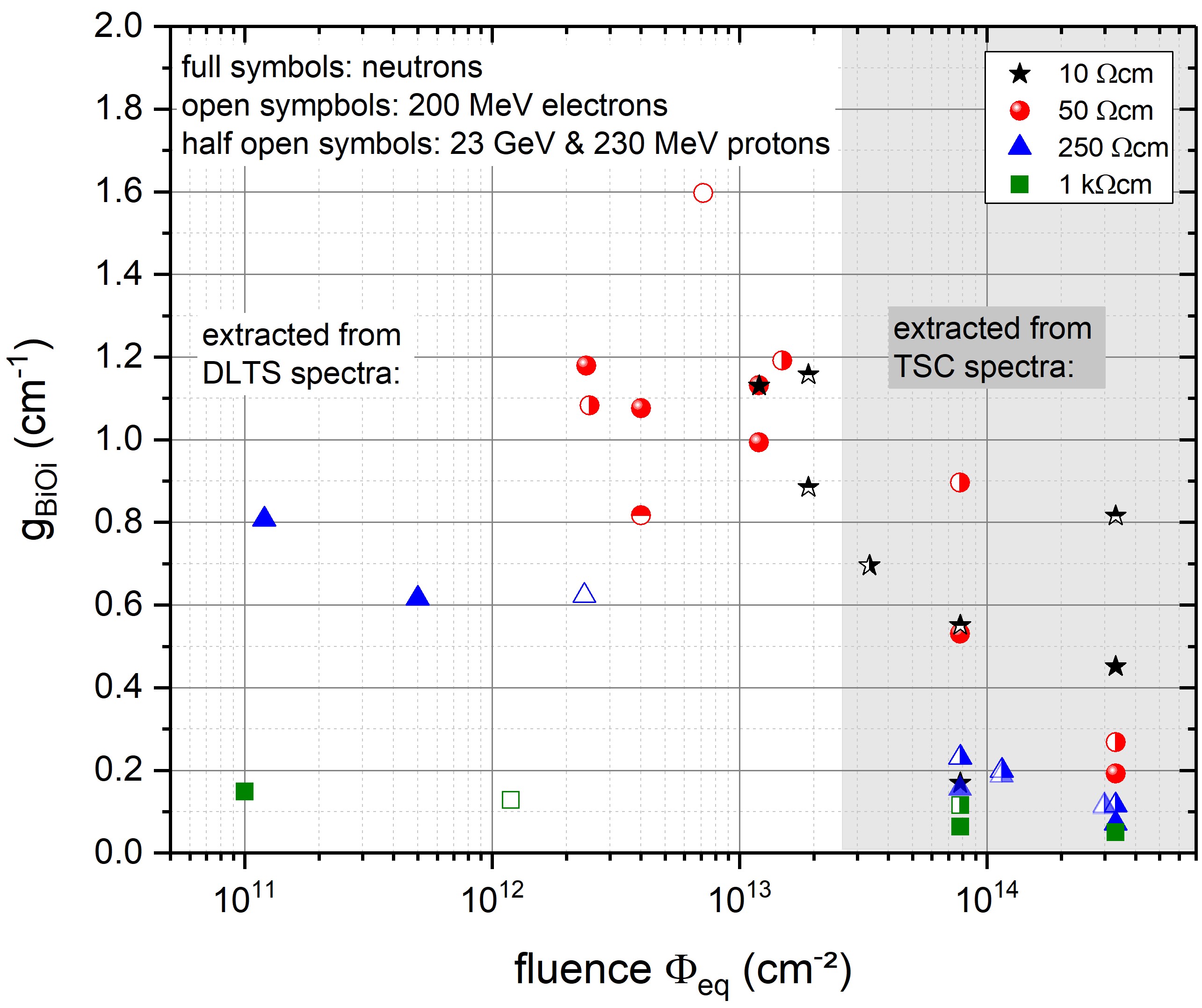}
    \caption{B$_\text{i}$O$_\text{i}$ introduction rates as function of the neutron equivalent fluence. The values below \num{3e13}\,\unit{n_{eq}\per\centi\square\meter} were calculated from B$_\text{i}$O$_\text{i}$ defect concentrations obtained from DLTS measurements on neutron (full symbols), electron (open symbols) and proton (half open symbols) irradiated EPI silicon pad diodes of different resistivity. For fluences above \num{3e13}\,\unit{n_{eq}\per\centi\square\meter} the data were extracted from the B$_\text{i}$O$_\text{i}$ peak measured by TSC (grey shaded area).}
    \label{fig:IR}
\end{figure}
%%%%%%%%%%%%%%%%%%%%%%%%%%%%%%%%%%%%%%%%%%%%%%%%
\section{Results and Discussion}
\label{sec:Results}
%%%%%%%%%%%%%%%%%%%%%%%%%%%%%%%
%%%%%%%%%%%%%%%%%%%%%%%%%%%%%%%
%%%%%%%%%%%%%%%%%%%%%%%%%%%%%%%
\subsection{Electrical characterization and applicability of the DLTS method to LGADs}

Before and after irradiation the LGADs and PIN diodes were electrically characterized by $C$-$V$ and $I$-$V$ measurements. The $I$-$V$ measurements showed an increase in the leakage current, while the current related damage factor $\alpha$ of \num{9.6e-19}\,A/cm, extracted from $I$-$V$ curves at -\,20$^{\circ}$C, agrees well with values given in the literature \cite{Moll-thesis}. Furthermore, after irradiation a shift of the break down voltage to higher values was observed, that however decreases with decreasing the measurement temperature.
$C$-$V$ measurements on LGADs after irradiation have shown a decrease in the gain layer capacitance as well as a decrease of the depletion voltage of the gain layer ($V_\text{GL}$) indicating the degradation of the gain layer due to the deactivation of active boron dopants. The $V_\text{GL}$ of the investigated LGADs as well as the depletion voltage of the corresponding PIN diodes are given in Table\,\ref{tab:samples}. 
%\todo[inline]{add the annealing status of the samples; add comment on the shift of VGL with fluence, is this in agreement with previous works? }
%%%%%%%%%% Figure %%%%%%%%%%%%%%%%%%%%%%%%%%%%%%
\begin{figure}[htb]
    \centering
    \includegraphics[width=1\columnwidth]{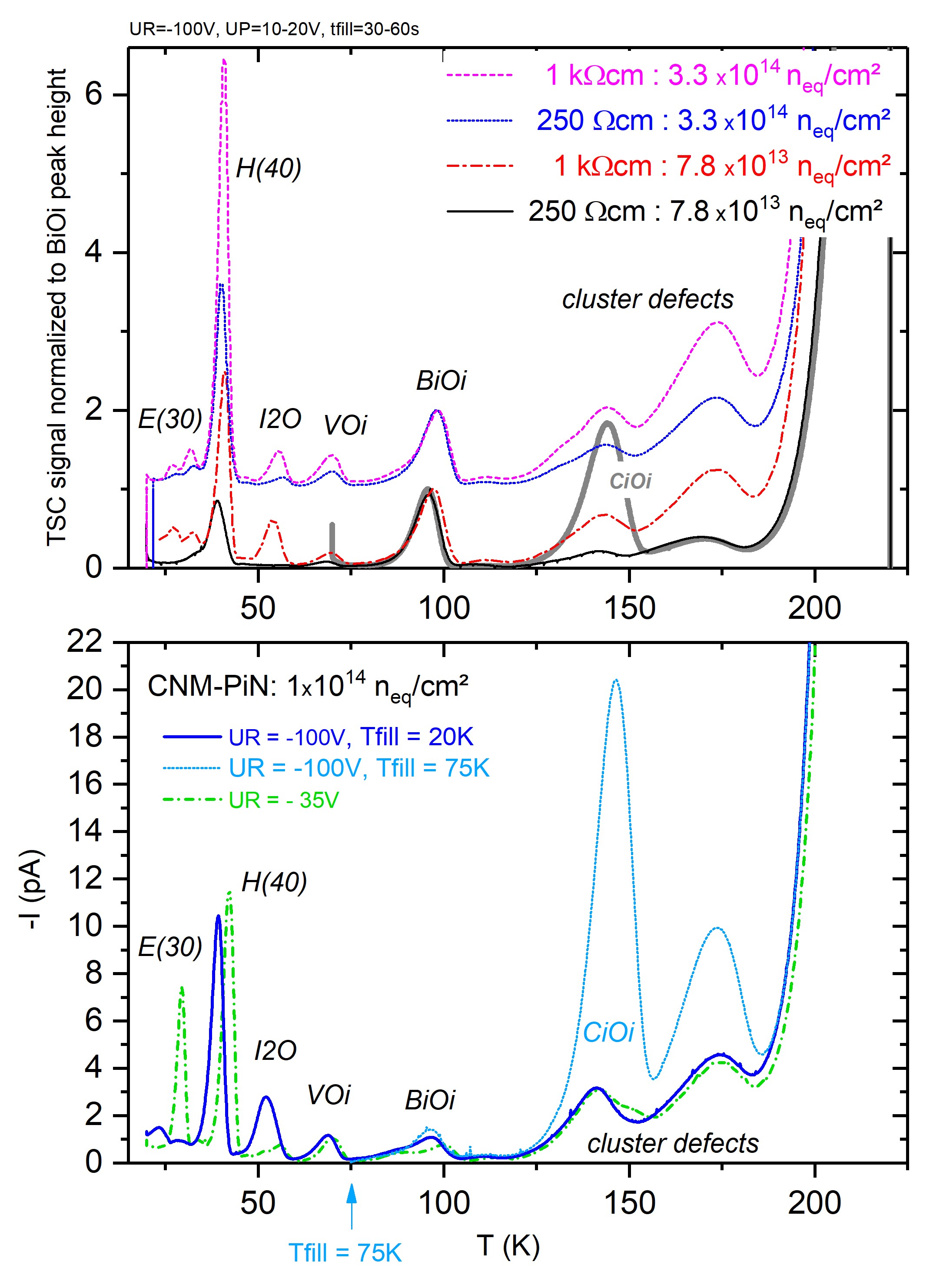}
    \caption{Top) TSC spectra of neutron irradiated EPI silicon pad diodes. The spectra are normalized to the B$_\text{i}$O$_\text{i}$ peak height and those with higher irradiation are shifted on the y-scale. The grey spectrum (bold solid line) was taken on a 250\,$\Omega$cm diode irradiated to \num{7.8e13}\,\unit{n_{eq}\per\centi\square\meter} and illustrates the filling temperature dependence of the C$_\text{i}$O$_\text{i}$ peak (\unit{T_{fill}}\,=\,70\,K). Bottom) TSC specta of a neutron irradiated CNM PIN diode. The spectra differ in the applied reverse bias. The light blue spectra (dotted line) was taken at UR\,=\,-100\,V and \unit{T_{fill}}\,=\,75\,K. [Coloured figures available online]}
    \label{fig:TSC-Pad-u-Pin}
\end{figure}
%%%%%%%%%%%%%%%%%%%%%%%%%%%%%%%%%%%%%%%%%%%%%%%%
Figure\,\ref{fig:CV} shows $C$-$V$ measurements performed at different measurement frequencies on a CNM LGAD neutron irradiated with \num{1e14}\,\unit{n_{eq}\per\centi\square\meter}. When using low measurement frequencies (standard: 10\,kHz) while increasing the reverse bias it can be well distinguished between the depletion of the gain layer region up to the gain layer depletion voltage ($V_\text{GL}$) and afterwards the depletion of LGAD bulk region. However for higher frequencies the measured capacitance of the gain layer region significantly drops. This effect becomes even more pronounced when decreasing the temperature as shown in Fig.\,\ref{fig:CF}. Here the capacitance values at a certain bias below $V_\text{GL}$ are plotted against the measurement frequencies. The data are taken in the temperature range of -\,20$^{\circ}$C to +\,20$^{\circ}$C and demonstrate the increased capacitance drop with decreasing temperature. 

To perform DLTS measurements, temperatures down to 20\,K were applied and a measurement frequency of \linebreak 1\,MHz is used. For the standard silicon pad diodes irradiated to adequate low fluence this allows the measurement of radiation induced defects very nicely as can be seen in Fig.\,\ref{fig:DLTS-EPI}. More details about the recorded spectra will be given below. On the other hand, DLTS measurements on LGADs 
resulted in non-reliable spectra due to the strong capacitance drop at high frequencies. Also the highly irradiated PIN diodes could not be measured with DLTS since in this high resistivity diodes the net background doping level is small compared to the high defect concentrations induced by irradiation. That leads to non-exponential capacitance transients during the thermal emission process and prevents reliable DLTS results \cite{Blood, Lang74}.   \\
The DLTS spectra in Fig.\,\ref{fig:DLTS-EPI} are measured on EPI silicon pad diodes of resistivities from 10\,$\Omega$cm to 1\,k$\Omega$cm. The neutron fluences are \num{1.3e13} and \num{1.2e13}\,\unit{n_{eq}\per\centi\square\meter} for the lower resistivity samples, and \num{1e11} and \num{5e11}\,\unit{n_{eq}\per\centi\square\meter} for the diodes with higher resistivity.
On the left-hand side, spectra obtained after injection of only majority carriers (holes) are shown while in the right-hand side the spectra after minority and majority carrier injection are plotted. As major electron trap the peak attributed to the B$_\text{i}$O$_\text{i}$ defect can be very nicely seen. Analysis of the capacitance transients gave the defect concentrations from which we extracted the defect introduction rate for B$_\text{i}$O$_\text{i}$. They are plotted in Fig.\,\ref{fig:IR} vs. particle fluence $\Phi_\text{eq}$ (white shaded area). The full symbols in this figure correspond to values taken from neutron irradiated diodes, while the open symbols and the half-open symbols are taken from comparable studies performed on 200\,MeV electron irradiated diodes and 23\,GeV as well as 230\,MeV proton irradiated EPI silicon pad diodes, respectively. 
As can be seen in Fig.\,\ref{fig:IR}, in the lower fluence range where DLTS is applicable the IR show rather a dependence on the device resistivity than on the fluence: for high resistivity material the IR are below 0.2\,cm$^{-1}$, medium resistivity samples show IR in the range of 0.6 to 0.8\,cm$^{-1}$, and for silicon pad diodes with low resistivities the values are above 0.8\,cm$^{-1}$ and in good agreement with previous experimental results on EPI silicon pad diodes \cite{Besleaga21NIMA, Liao2022IEEE} as well as with the defect kinetic model described in the introduction \cite{MOLL2019Vertex}. Furthermore, in this fluence range no clear dependence  of the IR on the particle type can be stated. However, a dependency of the IR on the particle type becomes visible at higher fluences, where the IR of proton irradiated sensors (half open symbols) are always higher than those of the neutron irradiated ones with the same resistivity (full symbols). This effect can be understood by a higher point defect formation ratio of protons compared to neutrons which preferentially create more cluster like defects \cite{Gurimskaya2020}. A more detailed discussion about the IR rates obtained by TSC (grey shaded area in Fig.\ref{fig:IR}) will be given in the next section.  \\
%%%%%%%%%%%%%%%%%%%%%%%%%%%%%%%%%%%%%%%%%%%%%%%%
%%%%%%%%%%%%%%%%%%%%%%%%%%%%%%%%%%%%%%%%%%%%%%%%
\subsection{TSC studies on silicon pad diodes and LGADs}
\begin{figure}[htb]
    \centering
    \includegraphics[width=1\columnwidth]{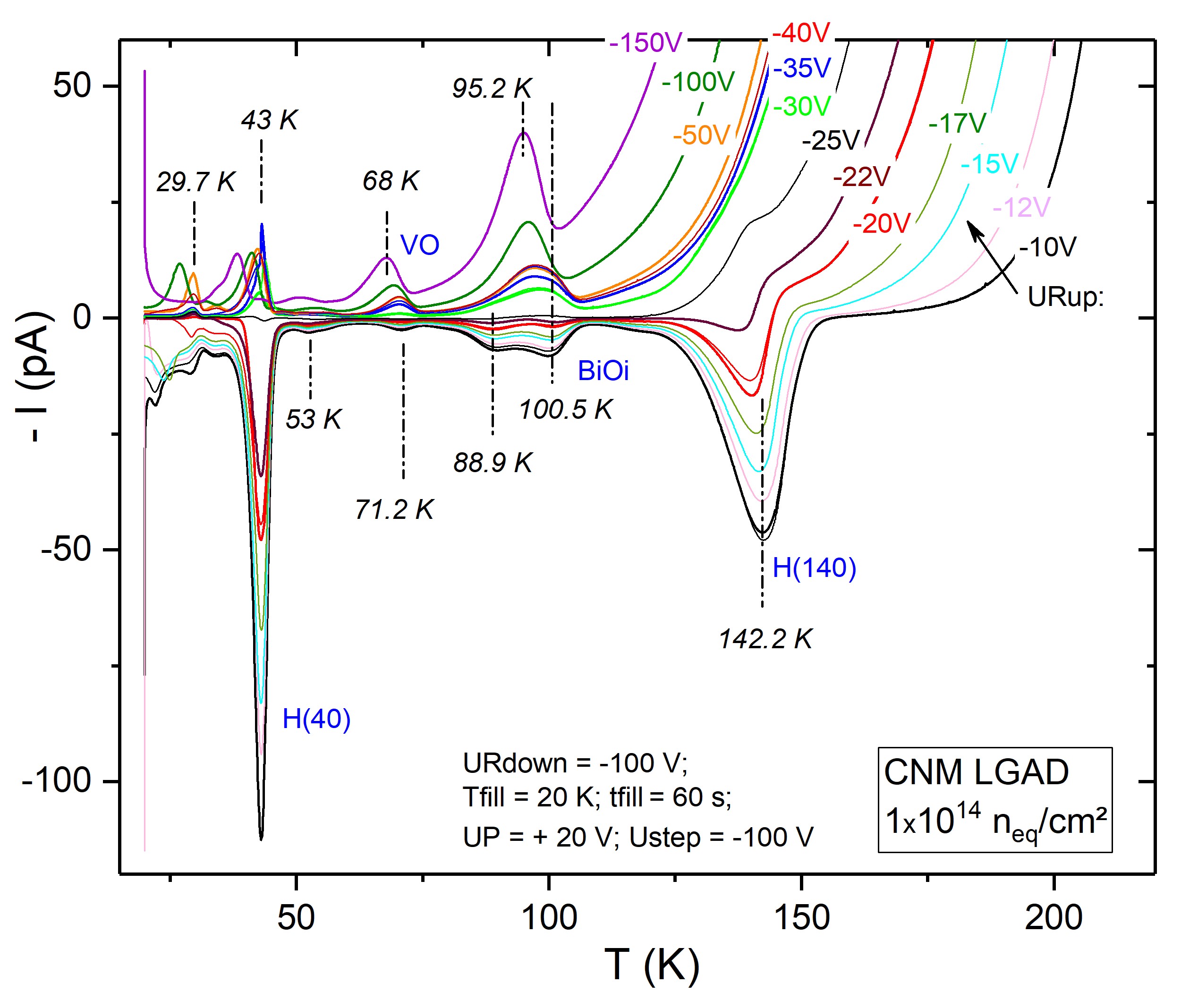}
    \caption{TSC spectra of a CNM LGAD irradiated with \num{1e14}\,cm$^{-2}$ neutrons. The spectra were measured using a "voltage step" (see corresponding description in the text) and applying different reverse bias during the "heating up" step. [Coloured figures available online] }
    \label{fig:TSC-voltage-step}
\end{figure}

Defect spectroscopy studies, including DLTS and TSC in combination with e.g. Electron Paramagnetic Resonance (EPR), Infra-Red absorption spectroscopy (IR) and detailed annealing studies, on a broad variety of irradiated silicon pad diodes have enabled the identification and assignment of a whole set of radiation induced defects in silicon \cite{Gurimskaya2020, Mooney1977PRB, Troxell1980, Fretwurst96NIMA, ZANGENBERG2002, Pintilie2006NIMA, Pintilie2008APL, Pintilie2009NIMA, Markevich2011PSSA, Radu2013NIMA, Radu2015JAP,Donegani2018NIMA}. Figure\,\ref{fig:TSC-Pad-u-Pin} (top) shows TSC spectra of silicon pad diodes with 250\,$\Omega$cm and 1\,k$\Omega$cm resistivity, neutron irradiated with \num{7.8e13}\,\unit{n_{eq}\per\centi\square\meter} and \num{3.3e14}\,\unit{n_{eq}\per\centi\square\meter}, respectively. The spectra are normalized to the B$_\text{i}$O$_\text{i}$ peak height and those of the higher irradiated diodes are shifted on the y-axis to allow better visibility. Besides point defects like E(30), H(40), I$_\text{2}$O, VO$_\text{i}$, B$_\text{i}$O$_\text{i}$ and C$_\text{i}$O$_\text{i}$, also multi-vacancy and cluster related defects (e.g. H(116), H(140), H(153)) \cite{Pintilie2009NIMA, Pintilie2008APL} at temperatures above 100\,K can be distinguished. Since the free carrier capture cross section of the C$_\text{i}$O$_\text{i}$ strongly depends on the filling temperature, it becomes visible in TSC only at higher \unit{T_{fill}} (see e.g. bold grey line in Fig.\,\ref{fig:TSC-Pad-u-Pin} with \unit{T_{fill}}\,=\,70\,K). In the temperature range of 96\,K\,--\,98\,K we can identify the B$_\text{i}$O$_\text{i}$ defect, that is assumed to be the main responsible defect for the radiation induced ARE.  \\
By fitting the TSC peak area and applying Equation\,\ref{eq:Nt} we extracted the B$_\text{i}$O$_\text{i}$ defect concentration and corresponding IR. The values are added in Fig.\,\ref{fig:IR}. All the IR for the B$_\text{i}$O$_\text{i}$ that were obtained by TSC for the diodes irradiated with neutrons (or protons) at higher fluences are below the values extracted from DLTS at lower fluences. Thereby it should be mentioned that for the highly doped diodes (10 and 50\,$\Omega$cm) an underestimation of the defect concentration has to be considered since those sensors were not able to fully deplete during the measurements. Therefore at least for those sensors the IR should be higher. The lower doped sensors (250 and 1\,k$\Omega$cm) can be fully depleted and show the same trend of a reduced IR. \\
When comparing the types of defects created in the EPI silicon pad diodes with those created in the CNM PIN (see Fig.\,\ref{fig:TSC-Pad-u-Pin} (bottom)) irradiated to \num{1e14}\,\unit{n_{eq}\per\centi\square\meter} it is found that they are very comparable. When changing the filling temperature also in the PIN the C$_\text{i}$O$_\text{i}$ peak becomes visible. The peak close to 100\,K can be assigned to the B$_\text{i}$O$_\text{i}$ with a concentration of about \num{1.5e12}\,cm$^{-3}$. To remember, the resistivity of the PIN diodes is below 5\,k$\Omega$cm, resulting in an effective doping level of about \num{2.6e12}\,cm$^{-3}$. The B$_\text{i}$O$_\text{i}$ IR would be around 0.02\,cm$^{-1}$. Here we should mention that the B$_\text{i}$O$_\text{i}$ peak shows kind of a shoulder at the lower temperature side whose origin is not fully clear, yet, but could be connected to the so-called X-defect presented in the literature \cite{Liao2022IEEE}. \\
When changing the applied reverse bias (from -100\,V to -35\,V) the influence of the electrical field to the carrier emission becomes visible, and effects especially the TSC peaks at low temperatures, corresponding to defect levels near to the band gap edges (E$_\text{C,V}$). This so-called Poole-Frenkel effect is explained by an enhanced emission probability from Coulombic traps due to a field induced lowering of the emission barrier height for trapped charges \cite{Frenkel38,Hartke68}. It leads in the TSC spectra to a peak shift to lower temperatures with increasing bias, and gives an explanation why for example the E(30) which is an electron trap with an energy level in the upper part of the band gap that shows a strong Poole-Frenkel effects \cite{Pintilie2009NIMA}, becomes visible only for low applied bias. \\
%%%%%%%%%% Figure %%%%%%%%%%%%%%%%%%%%%%%%%%%%%%
\begin{figure}[htb]
    \centering
    \includegraphics[width=1\columnwidth]{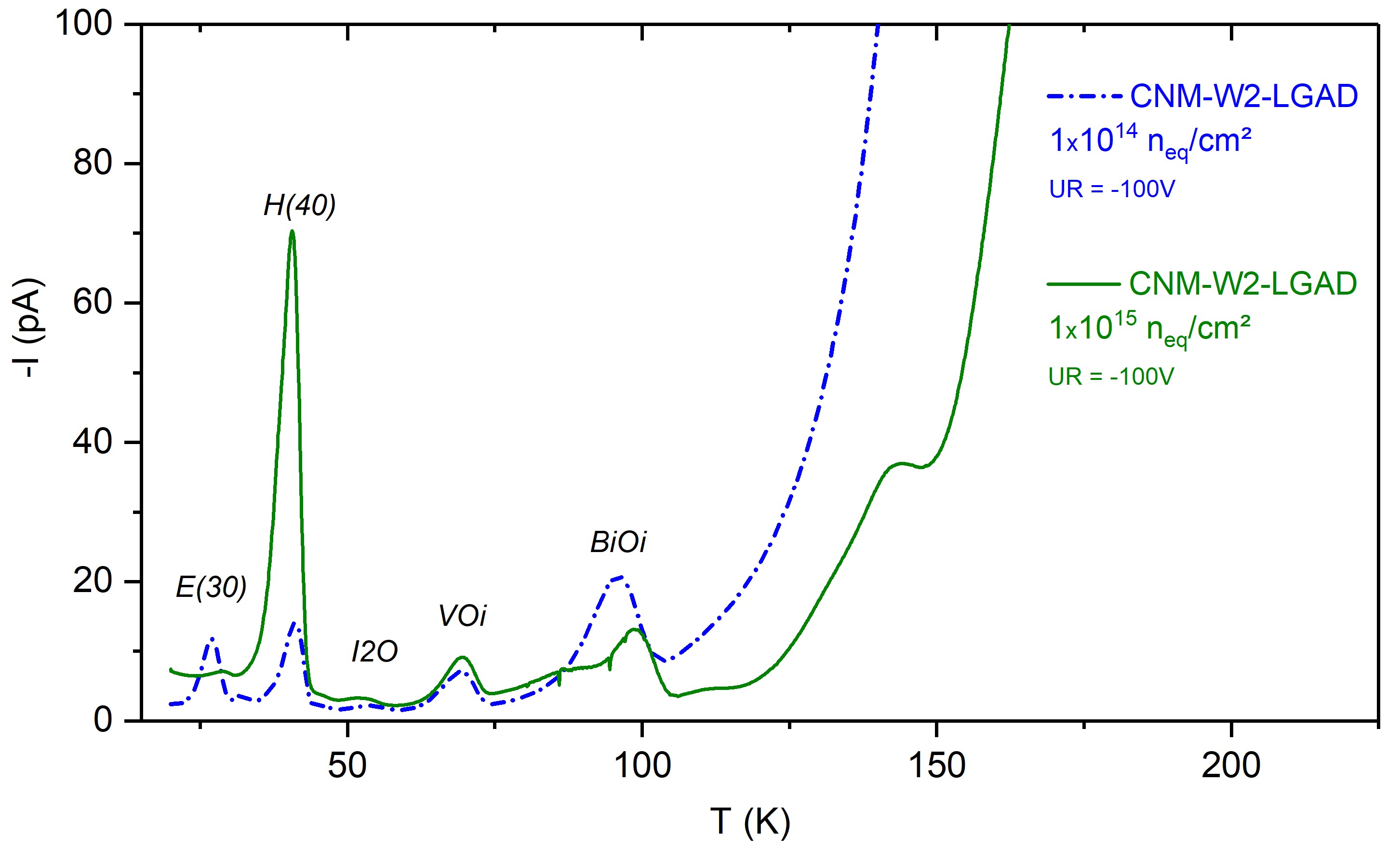}
    \caption{TSC spectra of LGADs neutron irradiated with different fluences. The TSC spectra were recorded while applying a reverse bias of -100\,V. [Coloured figures available online]}
    \label{fig:TSC-CNM-LGAD_HV}
\end{figure}
%%%%%%%%%%%%%%%%%%%%%%%%%%%%%%%%%%%%%%%%%%%%%%%%
%%%%%%%%%% Figure %%%%%%%%%%%%%%%%%%%%%%%%%%%%%%
\begin{figure}[htb]
    \centering
    \includegraphics[width=1\columnwidth]{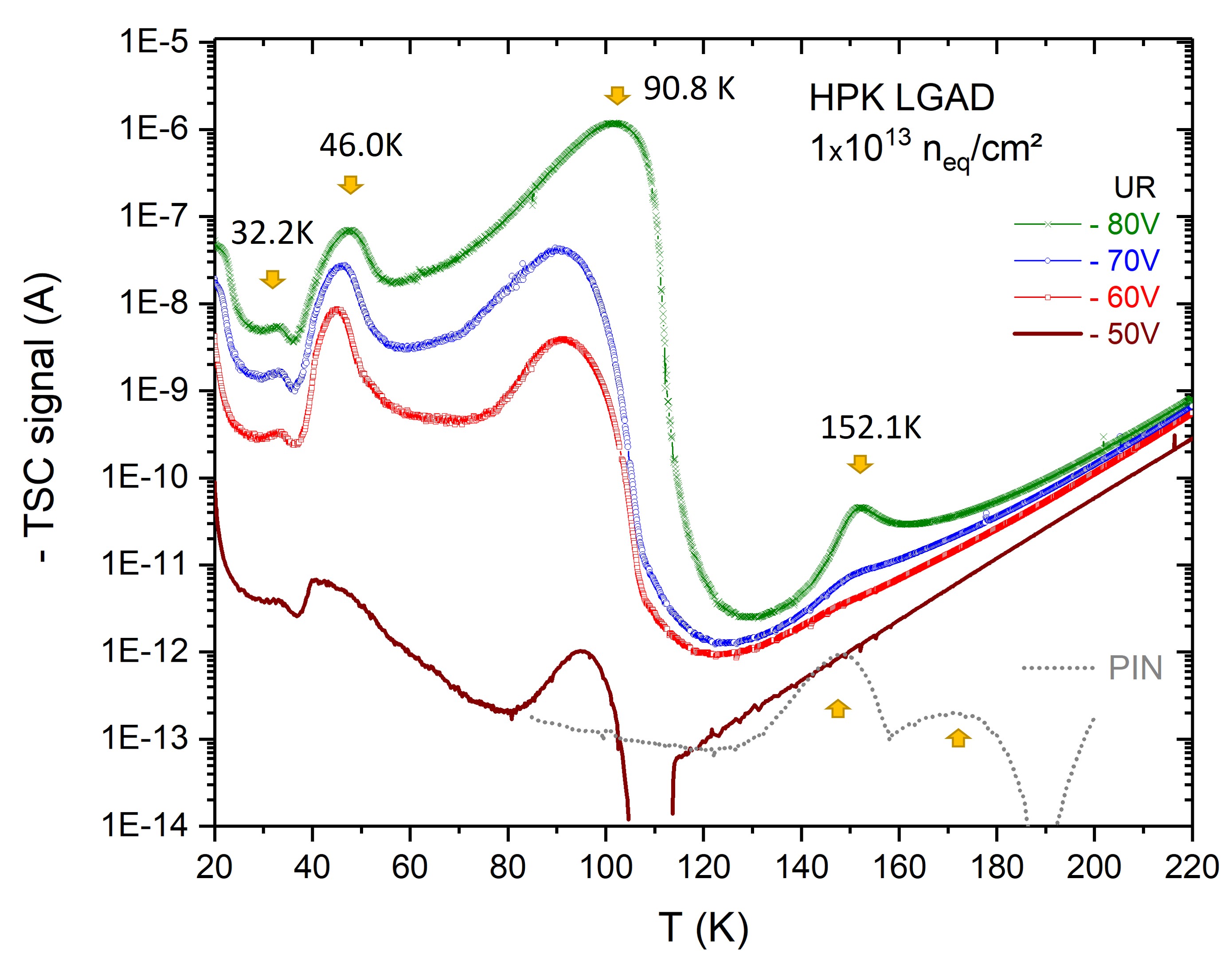}
    \caption{TSC spectra taken at different reverse bias of an LGAD irradiated with \num{1e13}\,\unit{n_{eq}\per\centi\square\meter} neutrons. The dotted grey line is the TSC spectrum of the corresponding PIN diode irradiated at the same fluence. [Coloured figures available online]}
    \label{fig:TSC-HPK_1E13}
\end{figure}
%%%%%%%%%%%%%%%%%%%%%%%%%%%%%%%%%%%%%%%%%%%%%%%%
\\
Figure\,\ref{fig:TSC-CNM-LGAD_HV} shows TSC spectra on CNM LGADs irradiated with different fluences (blue dash-dotted line:  \num{1e14}\,\unit{n_{eq}\per\centi\square\meter} and green solid line:  \num{1e15}\,\unit{n_{eq}\per\centi\square\meter}). The reverse bias applied during the measurement was in both cases U$_\text{R}$\,=\,-\,100\,V, corresponding to full depletion of the device (gain layer and low doped bulk). From these spectra we could identify the same defect types as measured in the PIN diode, however the peak heights are higher in the LGAD spectra. So the B$_\text{i}$O$_\text{i}$ concentration for the \num{1e14}\,\unit{n_{eq}\per\centi\square\meter} neutron irradiated LGAD is about \num{2E13}\,cm$^{-3}$ which gives an IR of about 0.2\,cm$^{-1}$. This IR is higher than for the PIN diodes but does not reach the values expected from the macroscopic sensor degradation given in the literature for highly doped LGAD gain layers \cite{MOLL2019Vertex}. Therefore, it is not very likely that this value reflects the deactivation of boron in the LGAD gain layer. 
Furthermore, it is observed that the increase of the background leakage current, contributing to the TSC signal, starts at temperatures significantly lower compared to the PIN or silicon pad diodes. This increased current signal is an effect of the charge amplification in the gain layer of the LGADs. Therefore, in the higher irradiated LGAD, where the gain is expected to be more reduced due to the radiation induced acceptor removal, the impact of the background leakage current starts at higher temperatures than for the lower irradiated device.   \\
The influence of the current amplification on the TSC signal becomes even more obvious when having a look on the measurements of \num{1e13}\,\unit{n_{eq}\per\centi\square\meter} irradiated HPK LGAD (see Fig.\,\ref{fig:TSC-HPK_1E13}). When applying a reverse bias of U$_\text{R}$\,=\,-\,80\,V that corresponds to a full depletion of the device, the current signal induced by charge emission from defect states reach values of up to \num{1e-6}\,A. That would correspond to defect concentrations in the range of 10$^{19}$\,cm$^{-3}$, being an effect of the charge amplification in the gain layer. Additionally it has to be taken into account, that for devices with multiplication layer the break-down voltage decreases with decreasing temperature \cite{Massey06}. For a non-irradiated HPK2-W36 LGAD it is around -50\,V below 120\,K compared to approximately -220\,V at room temperature. In consequence it means that an exact determination of defect concentrations from the measured TSC spectra is not reliable unless the exact impact of the gain at the given temperatures on the emitted charges is known. \\
In a next step we checked if it is possible to distinguish between defects created in the gain layer and defects created in the low doped bulk of the devices by restricting the depletion region to the width of the gain layer region. To do so the reverse bias applied during the TSC measurements was reduced. This allows to deplete only a certain part of the device, like e.g. only the gain layer region. Thereby, the emitted charges during the heating up step are supposed to be released mainly from defects within those parts. For the lower irradiated HPK LGAD the changes in the TSC signal with decreasing reverse bias are shown in Fig.\,\ref{fig:TSC-HPK_1E13}. 
It can be seen that, up to the full depletion voltage of the gain layer (-\,50\,V), the defect related TSC current signal more and more decreases and is not anymore detectable for biases below the gain layer depletion voltage. In conclusion, for the low irradiated HPK LGAD it is not possible to give a trustable explanation whether the charges multiplicated at higher biases are coming from the bulk or gain layer region. 
\\
%%%%%%%%%%%%%%%%%%%%%%%%%%%%%%%%%%%%%%%%%%%%%%%%
%%%%%%%%%% Figure %%%%%%%%%%%%%%%%%%%%%%%%%%%%%%
\begin{figure}[tb]
    \centering
    \includegraphics[width=1\columnwidth]{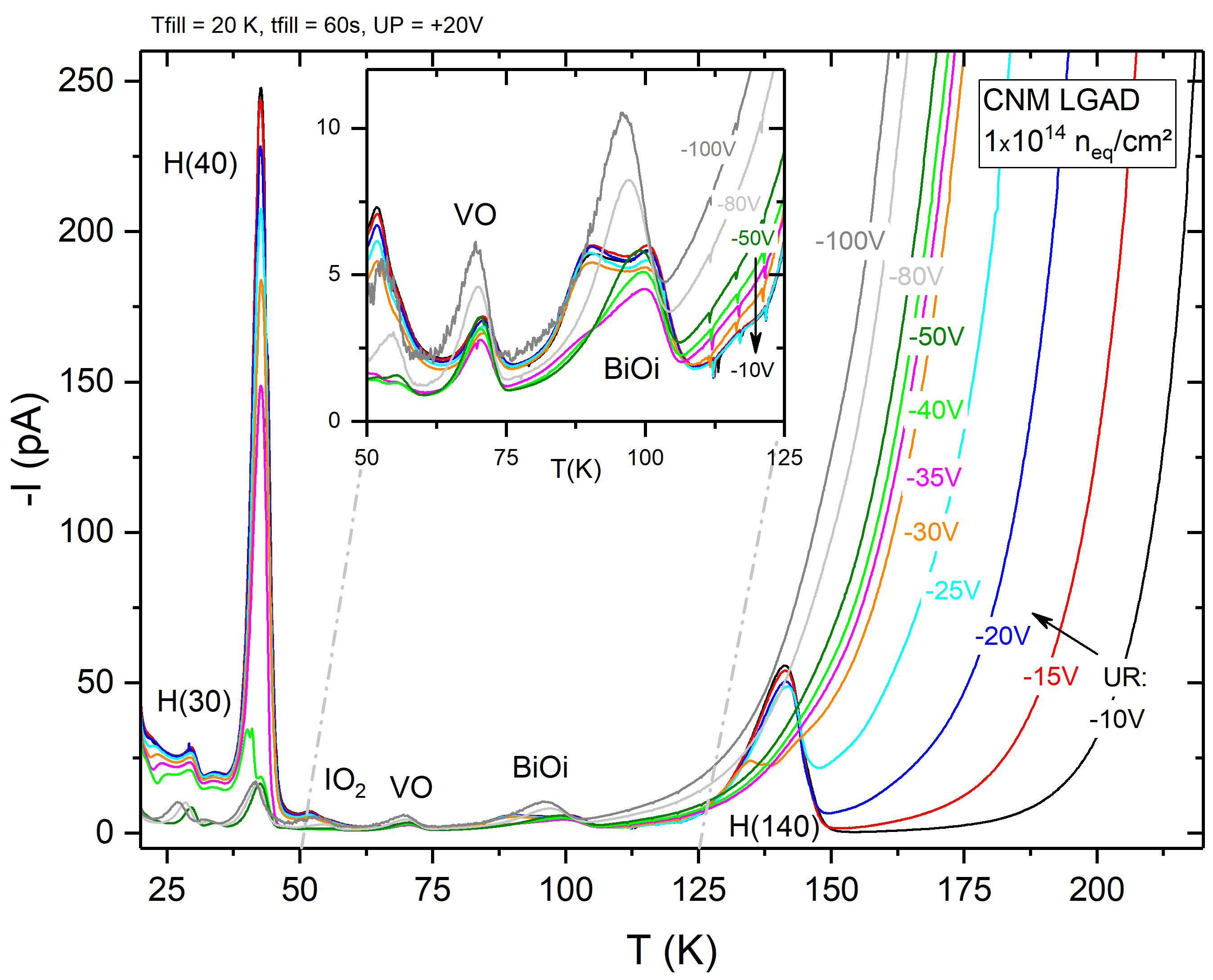}
    \caption{TSC spectra of a CNM LGAD irradiated with \num{1e14}\,cm$^2$ neutrons. The reverse bias voltage was varied in the range of -\,10\,V up to -\,100\,V. The inset shows a zoom of the spectra between 50\,K and 125\,K. [Coloured figures available online]}
    \label{fig:TSC-LGAD-diff-bias}
\end{figure}
%%%%%%%%%%%%%%%%%%%%%%%%%%%%%%%%%%%%%%%%%%%%%%%%
%%%%%%%%%% Figure %%%%%%%%%%%%%%%%%%%%%%%%%%%%%%
\begin{figure}[h!]
    \centering
    \includegraphics[width=1\columnwidth]{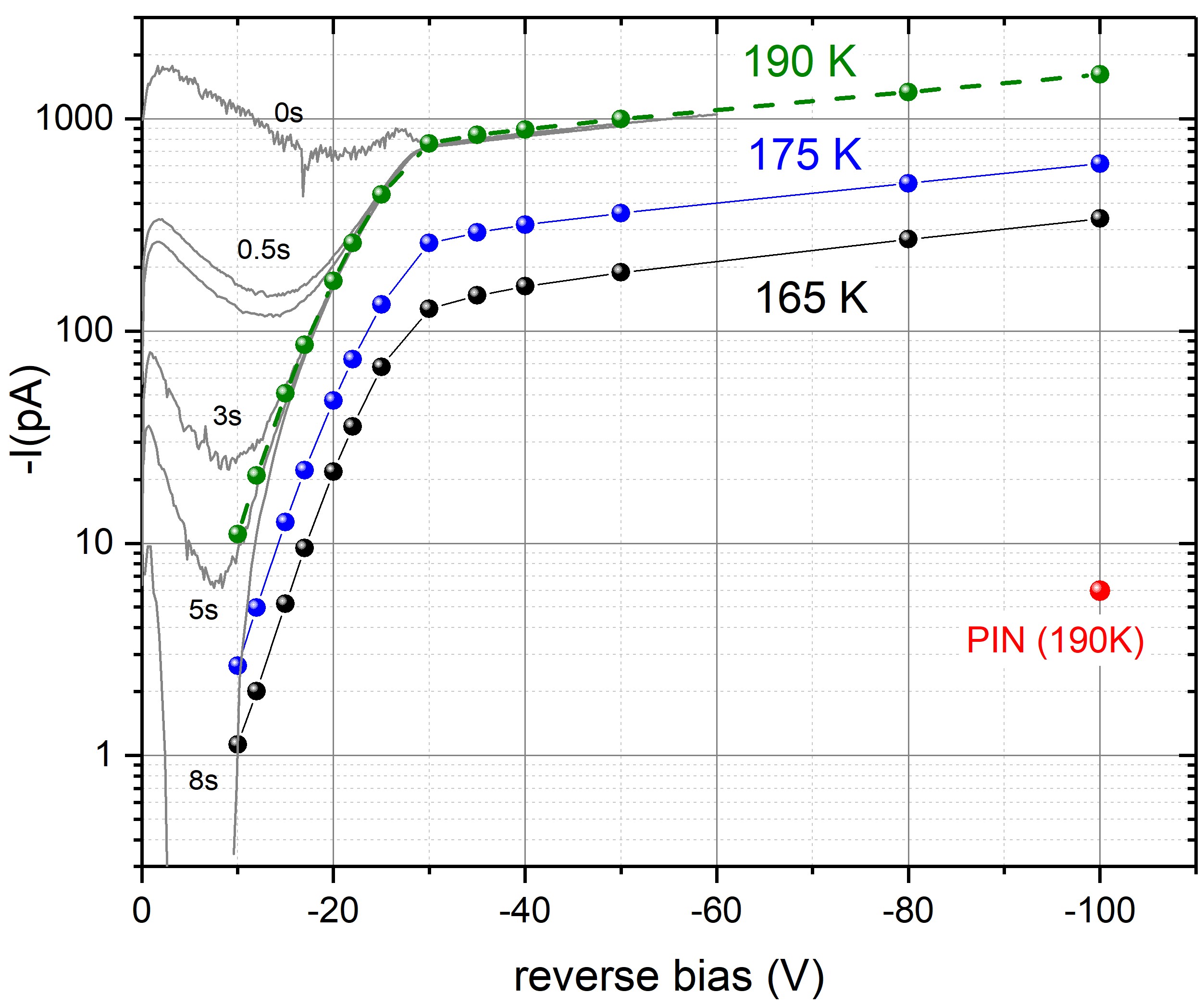}
    \caption{TSC current values extracted from the measurements plotted in Fig.\,\ref{fig:TSC-LGAD-diff-bias} at three different temperatures in dependency of the bias applied during the TSC measurement. Additionally added (solid grey lines): $I$-$V$ measurements on the same LGAD performed at 190\,K using different waiting times between two measurement points. Red dot labeled with PIN(190\,K): From TSC measurements extracted value of the corresponding PIN. [Coloured figures available online]}
    \label{fig:IV-temp}
\end{figure}
%%%%%%%%%%%%%%%%%%%%%%%%%%%%%%%%%%%%%%%%%%%%%%%%
%%%%%%%%%%%%%%%%%%%%%%%%%%%%%%%%%%%%%%%%%%%%%%%%
Comparable measurements performed on the higher irradiated CNM LGADs are illustrated in Fig.\,\ref{fig:TSC-LGAD-diff-bias}. First of all, one observes that with decreasing the bias voltage the background leakage current in the temperature region higher than 100\,\unit{\kelvin} decreases. The $I$-$V$ dependence for three different temperatures (165\,K, 175\,K and 190\,K), as extracted during TSC scan, is plotted in Fig.\,\ref{fig:IV-temp}. The red dot corresponds to the TSC current signal at 190\,K and -100\,V bias for the corresponding PIN diode. Additionally added are also $I$-$V$ measurements of the LGAD directly performed at 190\,K from low to high bias voltage. They were recorded with different waiting times between two measurement points ($\leq$\,10\,s). The longer the waiting time, the better is the agreement of the $I$-$V$ curve with the extracted values from TSC. The observed behaviour can be explained by taking into account the contribution of a time dependent displacement current at low bias. It even makes visible a sign inversion of the current signal for waiting times up to 8\,s leading to a minimum current value of -\,5.7\,pA. For waiting times higher than 8\,s no further changes in the measured $I$-$V$ curves were observed. In summary, the $I$-$V$ measurements reflect the behaviour observed during the TSC measurements and clearly demonstrates that at high voltages in the temperature range higher than 100\,K the TSC signal is dominated by a high background leakage current. Therefore, when lowering the voltages during the TSC scan defect levels like the H(140) becomes visible (see Fig.\,\ref{fig:TSC-LGAD-diff-bias}). But also the TSC current intensities of the peaks at lower emission temperatures are influenced by the applied voltages. Especially the defect levels detected below 50\,K strongly decrease in intensity when going from low to high voltages. \\
The inset in Fig.\,\ref{fig:TSC-LGAD-diff-bias} shows a zoom of the TSC spectra between 50\,K and 125\,K. Normally the B$_\text{i}$O$_\text{i}$ peak is expected to be measured at temperatures close to 100\,K. While for low voltages this peak consists of two maxima at higher voltages they merge into one peak. A double peak structure in this temperature range was also observed for the low doped PIN diodes and might be induced by the appearance of the X-defect \cite{Liao2022IEEE}.\\
Additionally to the current sign inversion that was visible in the $I$-$V$ measurements at low temperatures for low applied bias, also in TSC under certain measurement conditions an inversion in the TSC signal current was observed. This is illustrated in Fig.\,\ref{fig:TSC-voltage-step}. The TSC spectra here were recorded after cooling down with -100\,V reverse bias (TSC procedure: step 1), making a standard filling pulse (step 2) and after the filling pulse the reverse bias was set back to \unit{U_{step}}\,=\,-100\,V (in the following called "voltage step") before applying the reverse bias \unit{UR_{up}} for ramping up the temperature (step 3). As seen in Fig.\,\ref{fig:TSC-voltage-step}, if after the "high voltage step" the reverse bias for ramping up is chosen to be smaller than the gain layer depletion voltage (-\,10\,V to -\,25\,V), the current induced by the emitted charges from defect states has a positive sign. With increasing the bias voltage up to the gain layer depletion voltage the detected emission current decreases and increases again with opposite sign for bias higher than the gain layer depletion voltage (-\,30\,V to -\,150\,V). The observed shifts in the defect peak maxima can by understood by the Poole-Frenkel effect. The same behaviour is also observed for the lower irradiated HPK LGADs (not shown) and appears independent of the bias voltage applied during the TSC cooling down step. The high voltage step after the filling leads to a full depletion of the device. Putting afterwards the bias back to lower values most probably creates internal electric fields that counteract to the external applied field, leading charge carriers moving opposite to the external applied field. From literature it is known, that changes in the TSC current sign can appear due to internal residual electrical fields that are induced by high defect concentrations \cite{Pintilie2000NIMA, Bruzzi2009PoS,Bruzzi2010NIMA}.    
%%%%%%%%%% Figure %%%%%%%%%%%%%%%%%%%%%%%%%%%%%%
%%%%%%%%%%%%%%%%%%%%%%%%%%%%%%%%%%%%%%%%%%%%%%%%

\section{Summary}
In this paper we present defect spectroscopy studies (TSC and DLTS) on neutron irradiated silicon pad diodes as well as LGADs in order to understand the feasibility of using these methods to characterize defects in LGAD structures and find indications why the acceptor deactivation in the LGAD gain layer does not correlate with the defect kinetic model which is nicely applicable for the lower doped silicon pad diodes.
We demonstrate a strong dependency of the gain layer capacitance on the measurement frequency and temperature, resulting in a capacitance drop during DLTS measurements that makes this method not reliable for defect characterization of LGADs. In contrast, with TSC the defects formed in the LGAD devices can be very nicely detected and compared to studies performed on silicon pad diodes. However, giving quantitative values for the defect concentrations, and therefore exact introduction rates for the boron deactivation, is challenging since the peak amplitudes are determined by the multiplication effect of the gain layer and makes it in addition difficult to clearly distinguish between defect signals coming from the low doped Si bulk or from the highly doped Si gain layer. In conclusion, based on the experimental results it can be stated that the defect spectroscopy methods we used are limited in characterizing defects in LGAD gain layers. In order to use the full potential of these methods it is therefore planned in a next step to process and characterize highly-irradiated silicon pad diodes with doping concentrations that mimic the gain layer of an LGAD. 
%%%%%%%%%%%%%%%%%%%%%%%%%
\section*{CRediT authorship contribution statement}
\textbf{Anja Himmerlich:} Conceptualization, Investigation, Formal analysis, Validation, Writing - Original Draft 
\textbf{Nuria Castello-Mor:} Software, Formal analysis
\textbf{Esteban Curr\'{a}s Rivera:} Investigation, Formal analysis, Writing - Review \& Editing
\textbf{Yana Gurimskaya:} Investigation, Formal analysis, Validation, Writing - Review \& Editing
\textbf{Vendula Maulerova-Subert:} Software
\textbf{Michael Moll:} Conceptualization, Writing - Review \& Editing, Supervision, Project administration, Funding acquisition 
\textbf{Ioana Pintilie:} Validation, Conceptualization, Writing - Review \& Editing, Project administration, Funding acquisition
\textbf{Eckhart Fretwurst:} Conceptualization, Writing - Review \& Editing, Project administration, Funding acquisition
\textbf{Chuan Liao:} Validation, Writing - Review \& Editing 
\textbf{Jörn Schwandt:} Conceptualization, Writing - Review \& Editing, Project administration, Funding acquisition
%%%%%%%%%%%%%%%%%%%%%%%%%
\section*{Declaration of competing interests}
The authors declare that they have no known competing financial interests or personal relationships that could have appeared to influence the work reported in this paper.
%%%%%%%%%%%%%%%%%%%%%%%%%
\section*{Acknowledgement}
This work has been performed in the framework of the RD50 collaboration. I. Pintilie acknowledge the funding received 
through IFA-CERN-RO 08/2022 project. This project has received funding from the European Union’s Horizon 2020 Research and Innovation programme under  GA no 101004761.



\bibliographystyle{elsarticle-hard} 
\bibliography{literature}{}




%\bibitem{}

%\end{thebibliography}
\end{document}